\documentclass[11pt]{revtex4}

\pdfoutput=1

\usepackage{slashed}
\usepackage{mathtools}
\usepackage{hyperref}
\usepackage{amsmath}
\usepackage{amssymb}
\usepackage{latexsym}
\usepackage{amsfonts}
\usepackage{epsfig}
\usepackage{psfrag}
\usepackage{graphicx}
\usepackage{setspace}
\usepackage{braket}
\usepackage{mathrsfs}
\def\hhref#1{\href{http://arxiv.org/abs/#1}{arXiv:#1}} 
\usepackage[mathscr]{euscript}
\usepackage[scr]{rsfso}
\usepackage{color}

\newcommand{\bea}{\begin{eqnarray}}
\newcommand{\ea}{\end{eqnarray}}
\newcommand{\eea}{\end{eqnarray}}

\begin{document}

\linespread{1}

\title{Stokes Phenomenon and Hawking Radiation}

\author{Cesim K. Dumlu}

\affiliation{Extreme Light Infrastructure-Nuclear Physics (ELI-NP), 077125, M\u{a}gurele, Romania}

\email{cesim.dumlu@eli-np.ro}

\begin{abstract}
We compute the semiclassical decay rate for Kerr black hole by deriving a one-way connection formula, relating the near horizon solution to the outgoing solution at infinity. In particular, we discuss the relevance of the Stokes phenomenon and show how it leads to a Boltzmann-like thermal weight factor by  making use of the Stokes diagrams. We also give the exact result for the semiclassical greybody factor $e^{-2S}, $ where $S$ is  the leading order WKB action. We contrast our results with Maldacena and Strominger's work [Phys.\ Rev.\  D {\bf 56}, 4975 (1997)], where the emission spectrum for a rotating black hole was computed locally via asymptotic matching. We find that the relative error of semiclassical decay rate with respect to asymptotic matching formula diminishes in the limit of  large angular momentum, $l$, as expected. In this limit, the action assumes a compact form: $2S \sim (2 l+1) (\text{Log} \left( \frac{16 }{z_0}\right)-1)$, where $z_0$ is the cross ratio formed by the critical points (zeroes) of the scattering potential. 
\end{abstract}


\maketitle

\section{Introduction}

Since Hawking's original treatment of collapse geometry, the low energy decay rates for black holes has been a subject of extensive study\cite{hawk}. The thermal character of the emitted radiation discovered by Hawking was later derived within an alternative framework by Damour and Ruffini, who have considered analytic continuation of the outgoing modes in a fixed Kerr-Newman geometry\cite{dr}. Here, in the spirit of \cite{dr}, we consider an alternative approach for deriving Boltzmann-like thermal factor in Kerr background by incorporating the Stokes phenomenon.  As we will discuss in Section III in detail, we will derive a one way connection formula relating the amplitude of ingoing wave in the vicinity of the event horizon to the outgoing wave at infinity. The connection formula we obtain is derived from the standard analytic continuation rules of the semiclassical approximation. In this framework, obtaining the greybody factor is quite straightforward, the derivation of the Boltzmann-like factor on the other hand requires a careful analysis of the Stokes diagram in the vicinity of the event horizon. Our approach yields equivalent results with \cite{dr} without invoking negative norm states.

In recent years, the remarkable correspondence between the microscopic decay rate and the connection formula coming from the method of matched asymptotic expansions has received tremendous attention as one of the key aspects of  Kerr/CFT correspondence \cite{ms,cl,kcl1,kcl2,kcl3}.  In the asymptotic matching procedure, the subdominant terms in the scattering potential are either neglected or taken to be constant in near horizon and far regions. This enables the exact treatment of the radial problem in each respective region and because of this the accuracy of the emission rates  is expected to be very robust at low energies. Despite its accuracy, such an analysis is \textit{local} in nature as the scattering potential is treated separately in near and far regions. As a result, the greybody factors obtained in this fashion do not show explicit dependence on what may be called as the global features of the full scattering potential such as the cross ratios.  On the other hand, the exact treatment of the connection problem largely remains elusive. This is because in Schwarzschild/Kerr backgrounds the solutions of the radial problem are given by the Heun type functions, whose asymptotic expansions in terms of the known functions is less well known\cite{heun1,heun2}.  

In view of these observations  we analyze here the exact form of the leading order semiclassical action $S$ in Kerr geometry, to flesh out the analytical structure of the greybody factor which is otherwise uncaptured by the asymptotic matching procedure.  In contrast to the asymptotic matching procedure, WKB solutions for the  radial problem are approximate but no approximation on the scattering potential is made. Due to approximate nature of solutions,  the accuracy of WKB connection formula is limited, giving  accurate results for higher partial waves, but the semiclassical action $S$ reveals an interesting structure. Being invariant under translations, $S$ is given in terms of elliptic integrals whose arguments depend on three independent cross ratios, $z_0, z_{\pm}$ formed by the critical points (zeroes) and the poles (event horizons) of the scattering potential.

The plan of the paper is as follows. In Section II, we briefly recall the asymptotic matching procedure in Kerr background. In Section III, we give the analytic continuation rules and derive the connection formula in accordance with the Stokes diagram of WKB solutions. We discuss the analytic properties of $S$ in section IV. We particularly look into the limit where $z_0 \ll 1$ and we also give quantitative results on the relative error of the connection formula with respect to the asymptotic matching formula given by Maldacena and Strominger \cite{ms}. The final section contains our conclusions.

\section{Asymptotic Matching}

In this section we briefly recall asymptotic matching procedure,  closely following the prescription given in \cite{ms}.  The background we will work with is given by the Kerr metric:
\bea
ds^2 &=&-\left(\frac{\Delta- a^2 \sin{\theta}^2}{\Sigma}\right) d t^2 + \frac{\Sigma}{\Delta} \,dr^2 +\Sigma \,d\theta^2  +\left(\frac{(r^2+a^2)^2- \Delta a^2 \sin^2{\theta}}{\Sigma}\right) \sin^2{\theta} \, d\phi^2 \nonumber\\
&-&  \left(\frac{2 a \sin^2{\theta} (r^2+a^2-\Delta)}{\Sigma}\right) d t \,d\phi
\eea
where
\bea
\Sigma=r^2+a^2 \cos^2{\theta}, \quad
\Delta= r^2 -2 M r +a^2
\eea
We will use normalized units throughout ($G=c=1$). The event horizons are given by the roots of $\Delta$:
\bea
r_{\pm} = M \pm \sqrt{M^2-a^2}
\eea
The remaining relevant quantities, the temperature $T_{H}$, angular velocity $\Omega$, and area $\mathcal{A}$ of the horizon are defined as:
\bea
T_{H}=\frac{r_+ - r_-}{8 \pi M r_+}, \quad  \Omega=\frac{a}{2 M r_+}, \quad \mathcal{A}=8\pi M r_+
\eea
We will consider the matching problem for the radial component of a  massless scalar field satisfying the wave equation:
\bea
\frac{1}{\sqrt{-g}}\partial_{\mu} \left(\sqrt{-g}\,g^{\mu\nu} \partial_{\nu} \Phi \right)=0
\label{scad}
\eea 
which admits separable solution of the form:
\bea
\Phi=e^{i m \phi - i \omega t}S_{\lambda} (\theta, \, a\,\omega) \, R(r)
\label{sca}
\eea
By plugging (\ref{sca}) into (\ref{scad}), one gets for the radial part: 
\bea
\Delta\frac{d}{dr} \left(\Delta  \frac{d R}{d r} \right) + \left( K^2 -\Lambda \Delta\right) R=0
\label{radiald}
\eea 
where:
\bea
K=(r^2 + a^2)\,\omega - a m , \quad \Lambda=\lambda + a^2 \omega^2 - 2 a m \omega
\eea
For $a \,\omega \ll 1$, the eigenvalues satisfy: $\lambda \approx \ell (\ell+1) + \mathcal{O}( a^2 \omega^2 )$.  In what follows the exact solutions of (\ref{radiald}) in near and far horizon regions will be given. These solutions will be matched in the overlapping region, giving us the matching coefficient that determines the emission amplitude. The precise definitions of near and far regions were given in \cite{ms} as:
\bea
\text{Near region} &:&  \omega (r-r_+) \ll  1 \nonumber\\
\text{Far region} &:&   M \ll r-r_+
\label{match}
\eea   
Note that the conditions given above makes no reference to angular momentum, $\ell$. For large $\ell$, there is a parameter window for which above matching conditions can be loosened. This was elucidated by Cvetic and Larsen in \cite{cl} where a consistent set of matching conditions involving the eigenvalues were given.  The low energy condition is given by $M \omega \ll 1$, which means that Compton wavelength of the scalar particle is much bigger than the gravitational size of black hole. From the WKB perspective,  the low energy limit or more precisely the correspondence limit translates into the fact that there is a large separation between the critical points (zeroes) of the scattering potential for the given value of momenta. This geometric notion is encoded by the condition: $z_0 \ll 1$,  where $z_0$ is the cross ratio formed by the zeroes of the scattering potential. We will come back to this point later in section III.   

\subsection{Near region}

Using (\ref{match}), the scattering potential in the near region is approximated as:
\bea
K^2- \Lambda \Delta \approx r_+^{4}(\omega - m \Omega)^2 - \ell (\ell+1) \Delta
\eea
where $a^2\omega$ and smaller terms are neglected. With these approximations the radial equation takes the form:
\bea
\Delta \frac{d}{dr} \left(\Delta  \frac{d R}{d r} \right) + \left( r_+^{4}(\omega - m \Omega)^2 - \ell (\ell+1) \Delta\right) R=0
\label{radialdn}
\eea 
which upon making the following substitutions:
\bea
z=\frac{r-r_+}{r-r_-}, \quad R=A \, z^{i \frac{\omega- m \Omega}{4\pi T_{H}}} (1-z)^{\ell+1} F
\label{subs}
\eea
turns into the hypergeometric differential equation:
\bea
z (1-z) \partial^2_z F + \left( \gamma - (1+ \alpha + \beta) \right) \partial_z F - \alpha \beta F
\label{hyperd}
\eea
with:
\bea
F&:=& F(\alpha, \, \beta, \, \gamma\, ; z)\nonumber\\
\alpha &=& 1+ \ell + i \frac{\omega- m \Omega}{2\pi T_{H}}, \nonumber\\
\beta   &=&  1+ \ell \nonumber\\
\gamma &=& 1+ i \frac{\omega- m \Omega}{2 \pi T_H}
\label{hypers}
\eea
Note that (\ref{hyperd}) has two independent solutions. The solution in (\ref{hypers}) was picked on the grounds that there should be only ingoing wave at the horizon, $z=0$. (we will elaborate on the boundary conditions in more detail when we discuss the Stokes phenomenon) This solution is to be matched with the far region solution and for this we need the asymptotic expansion of (\ref{hypers}) for large $r$ ($z\rightarrow 1$). The asymptotic expansion follows from the Kummer transformation of the hypergeometric function yielding:
\bea
 F(\alpha, \, \beta, \, \gamma\, ; z) &=& \left(1-z \right)^{\gamma-\alpha-\beta} \frac{\Gamma(\gamma)\Gamma(\alpha+\beta-\gamma)}{\Gamma(\alpha)\Gamma(\beta)} F(\gamma-\alpha, \, \gamma-\beta, \, \gamma-\alpha-\beta+1\, ;1- z) \nonumber\\ &+& \frac{\Gamma(\gamma)\Gamma(\gamma-\alpha-\beta)}{\Gamma(\gamma-\alpha)\Gamma(\gamma-\beta)} F(\alpha, \, \beta, \, \alpha+\beta-\gamma+1\, ;1- z) 
\eea   
Letting $z$ to unity and using (\ref{subs}) ultimately yields:
\bea
R&=&A \left(\frac{r} {r_+-r_-}\right)^{-\ell-1}\Gamma\left(1+ i \frac{\omega- m \Omega}{2 \pi T_H} \right) \times \nonumber\\
 && \left( \frac{\Gamma(-2\ell-1)}{\Gamma(-\ell)\Gamma\left(-\ell+ i \frac{\omega- m \Omega}{2 \pi T_H}\right)} +  \frac{\Gamma(2\ell+1)}{\Gamma(\ell+1)\Gamma\left(1+ \ell+  i \frac{\omega- m \Omega}{2 \pi T_H}\right)} \left(\frac{r} {r_+-r_-}\right)^{2\ell+1} \right)
\label{hyperm}
\eea

\subsection{Far region}
In this region gravitational effects die out, the radial equation reduces to the ordinary differential equation in flat space:
\bea
\frac{1}{r^2}\partial_r (r^2 \partial_r R) + \frac{\omega^2 r^2 -\ell (\ell+1)}{r^2}R=0
\eea
which admits solutions given by Bessel functions:
\bea
R=\frac{1}{\sqrt{r}}\left[a_1 J_{\ell+\frac{1}{2}} (\omega r) + a_2 J_{-\ell-\frac{1}{2}} (\omega r)   \right]
\label{bessel}
\eea
where $a_1$ and $a_2$ are unspecified coefficients. For large $r$ above solution behaves as:
\bea
R \xrightarrow{r \rightarrow \infty}  \frac{1}{r}\sqrt{\frac{2}{\pi \omega}}\left[-a_1 \sin{\left(\omega r- \frac{\ell \pi}{2} \right)}+a_2 \cos{\left(\omega r+ \frac{\ell \pi}{2} \right)}\right]
\label{besself}
\eea
We will use this solution to compute the outgoing flux at infinity. The form of the solution to be matched with (\ref{hyperm}) is given by  the small $r$ expansion of  (\ref{bessel}), which is approximately given by: 
\bea
R \simeq \frac{1}{\sqrt{r}}\left[ \frac{a_1}{\Gamma(\ell +\frac{3}{2})} \left(\frac{\omega r}{2}\right)^{\ell +\frac{1}{2}} +\frac{a_2}{\Gamma(-\ell +\frac{1}{2})} \left(\frac{\omega r}{2}\right)^{-\ell -\frac{1}{2}} \right]
\label{besselm}
\eea

\subsection{Matching $\&$ Decay Rate}

The remaining task to is to equate the dominant part of the solutions (with positive powers)  given by (\ref{hyperm}) and (\ref{besselm}). This specifies the coefficient $A$ to be:
\bea
A=\frac{\left(r_+-r_-\right)^{\ell} \omega^{\ell+\frac{1}{2}}\Gamma(\ell+1)\Gamma\left(1+\ell+ i \frac{\omega- m \Omega}{2 \pi T_H}\right)}{2^{\ell+\frac{1}{2}}\Gamma(\ell+\frac{3}{2})\Gamma(2\ell+1)\Gamma\left(1+ i \frac{\omega- m \Omega}{2 \pi T_H}\right)} a_1
\eea
To find the  decay rate, we need the ingoing flux at the horizon and the outgoing flux at infinity. These can be readily computed by using the  conserved flux for the radial equation (\ref{radiald}):
\bea
f=\frac{2\pi}{i}{R^{*}\Delta \partial_r R - R\Delta \partial_r R^{*}}
\eea
and the solutions (\ref{subs}) and (\ref{besself}), yielding (see \cite{ms} for details) :  
\bea
f_{\text{in}}= \frac{\omega-m\Omega}{T_{H}}(r_+-r_-)\left|A\right|^2, \quad f_{\text{out}}= 2 \left|a_1\right|^2
\label{flux}
\eea
The absorption amplitude is given by the ratio of fluxes:
\bea
\sigma=\frac{\omega-m\Omega}{2T_{H}} \, \frac{(r_+-r_-)^{2\ell+1}\omega^{2\ell+1}}{2^{2\ell+1}} \, \left| \frac{\Gamma(\ell+1)\Gamma\left(1+\ell+ i \frac{\omega- m \Omega}{2 \pi T_H}\right)}{\Gamma(\ell+\frac{3}{2})\Gamma(2\ell+1)\Gamma\left(1+ i \frac{\omega- m \Omega}{2 \pi T_H}\right)} \right|^2  
\label{abs}
\eea
and the total decay rate is:
\bea
\Gamma=\frac{\sigma}{e^{\frac{\omega-m\Omega}{T_{H}}}-1}
\label{decay}
\eea
The full expression for the decay rate can  further be simplified by making use of the identities for the gamma function, but for the sake of comparison with the WKB connection formula we will use the factorized form given above.   

\section{Semiclassical Approximation}
 \begin{figure*}
\includegraphics[width=9cm,height=6cm]{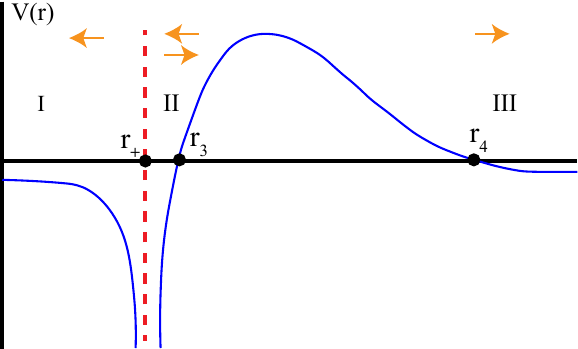}
\caption{Schematic representation of effective scattering potential $V(r)$ (solid blue line). The points $r_3$ and $r_4$ denote the zeroes of $V(r)$. The region I and II are respectively located to the left and to the right of the event horizon $r_+$, marked by the dashed vertical line. The arrows indicate the local form of the solution in each respective region: In region I, there is only ingoing solution, whereas region II contains both ingoing and outgoing solutions. The region III is located beyond the critical point $r_4$, where there is only outgoing solution.}
\label{f1}
\end{figure*}
In order to perform the semiclassical analysis we return back to radial equation in (\ref{radiald}), and make the substitution $R=\Delta^{-\frac{1}{2}}\tilde{R}$. This eliminates the first order derivative and turns (\ref{radiald}) into Schr\"odinger-like equation:
\bea
\partial^2_r\tilde{R} + \left(\frac{K^2-\Lambda \Delta + M^2-a^2}{\Delta^2}\right)\tilde{R}=0
\label{schro}
\eea
with the effective scattering potential:
\bea
V(r)=-\left(\frac{K^2-\Lambda \Delta + M^2-a^2}{\Delta^2}\right)
\eea
and admits the standard WKB solutions:
\bea
\tilde{R}=\dfrac{a_1}{\sqrt{2\, p_r(r)}}e^{\varphi(p,\, r)} + \dfrac{a_2}{\sqrt{2\, p_r(r)}}e^{\varphi(r,\, p)} 
\label{wkbs}
\eea
Here, $a_1$ and $a_2$ are undetermined complex coefficients.  The phase integral is defined as:
\bea
\varphi(p,\, r)= i\int^r_p p_r(r')\, dr'
\eea
with the limits of integration are interchanged for $\varphi(r,\, p)$. Note that the specific value of $\varphi(p,\, r)$ is defined with respect to the phase reference point $p$, which is generically taken to be the critical points of potential.  The radial momentum $p_r$ is given by the positive root:
\bea
p_r =\frac{\left[K^2-\Lambda \Delta + M^2-a^2\right]^{\frac{1}{2}}}{\Delta}
\label{radialm}
\eea 
such that: $\lim_{r\rightarrow \infty} p_r = \omega $. The appropriate boundary conditions for the problem can directly be read-off the scattering potential as shown in Fig.(\ref{f1}). In the horizon region we have over-the-barrier scattering. This means in the region I, to the left of the event horizon, we have only ingoing solution as required by causality, whereas in region II, in the vicinity of the critical point, there are ingoing and outgoing solutions.  The regions II and III are connected by under-the-barrier transmission between two critical points therefore we require there is only outgoing wave in region III.

\subsection{Analytic Continuation and the Stokes Phenomenon}

In the following,  we will briefly introduce the  analytic continuation rules and use these rules to derive a connection formula between ingoing modes in the vicinity of the horizon and the outgoing modes at infinity. For an in-depth analysis of the theory and the Stokes phenomenon, we refer the interested reader to an excellent review given by Berry\cite{berry} and the classic manuscript by Heading\cite{heading}.

\begin{figure*}
\includegraphics[width=17cm,height=8cm]{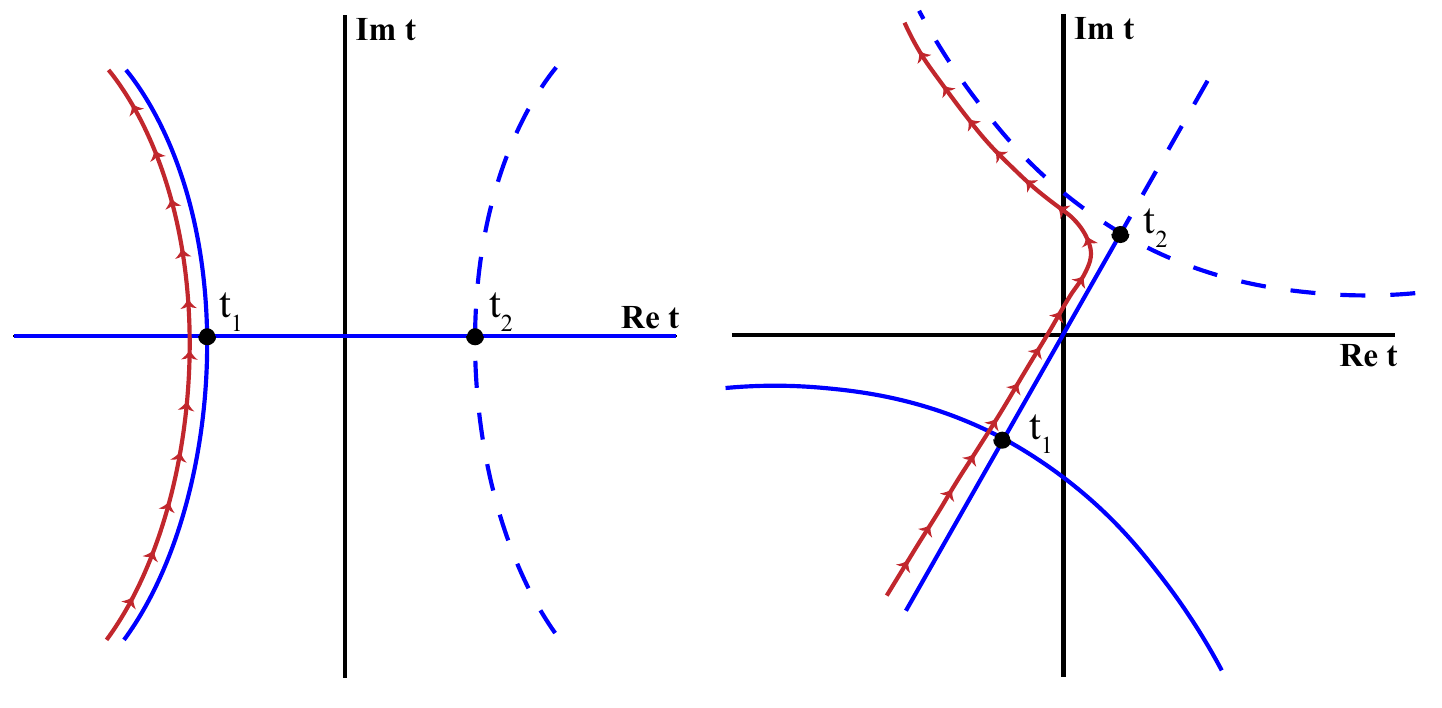}
\caption{Integration contours, the lines of steepest descent and ascent for the Airy function given by (\ref{airyI}). For $\text{Arg}(r)=0$ (left) the  saddle points are respectively located at $t_1=-r^{1/2}$ and $t_2=r^{1/2}$. The solid horizontal (blue) line represents the line steepest ascent through $t_1$ and the line of steepest descent through $t_2$. The left solid curve is the line of steepest descent through $t_1$, and it begins and ends where $\text{Arg}(t)\sim \pm\frac{2}{3}\pi$, which is within the sectors given by (\ref{sect}). Hence, the integration contour shown by arrowheads can be deformed to coincide with the line of steepest descent through $t_1$. For $\text{Arg}(r)=\frac{2}{3}\pi$ (right), the saddle  points shift  to the complex plane. The integration contour in this case is aligned with the line of steepest descent through $t_1$ all the way up to $t_2$. To continue the descent and end in the right sector, the contour makes a right angle in the vicinity of $t_2$ and continues on the left branch.   }
\label{f2}
\end{figure*}

The boundary conditions discussed in the previous section show that the form of the solution given by (\ref{wkbs}) is local in nature. In between critical points the solution is evanescent, in other words, there is only the exponentially decaying solution. If we pass into the region II, not only the solution becomes oscillatory but it also becomes a linear combination of both solutions.  In semiclassics, the appearance of the second solution upon changing the complex domain is known as the Stokes phenomenon.  A canonical example of this is given by the Airy function $\text{Ai}(r)$, whose behavior in the complex domain is analogous to that of WKB solutions here. For complex $r$, $\text{Ai}(r)$ has the following representations:
\bea
\text{Ai}(r) &\sim& \frac{1}{2} \pi^{-\frac{1}{2}} r^{-\frac{1}{4}} e^{-\frac{2}{3} r^{3/2}} , \quad \text{Arg}(r)=0 \nonumber\\
\text{Ai}(r) &\sim& \frac{1}{2} \pi^{-\frac{1}{2}} r^{-\frac{1}{4}} (e^{-\frac{2}{3} r^{3/2}} + i\, e^{\frac{2}{3} r^{3/2}} )\, , \quad  \frac{2\pi}{3} \leq \text{Arg}(r) \leq \frac{4\pi}{3} 
\label{airy}
\eea
Note that the function behaves exponentially small (subdominant) for real and positive $r$, i.e to the right of the critical point at $r=0$. However, this property reverses on the line emanating from the critical point, with $\text{Arg}(r)=\pi/3$. This line is called the anti-Stokes line on which the real part of the exponent vanishes and changes its sign. This simple observation immediately leads us to the first rule: upon crossing an anti-Stokes line,  a subdominant(exponentially small) solution turns into dominant (exponentially large) solution and vice versa. For WKB solutions, we can denote this change by making use of the compact notation introduced by Heading\cite{heading}:
\bea
\frac{1}{\sqrt{2\, p (r)}}e^{\varphi(p,\,r) } := \left(p,\, r\right), \quad  \frac{1}{\sqrt{2\, p (r)}}e^{\varphi(r,\,p) } := \left(r,\, p\right)
\eea  
Given the subdominant solution, $\left(p,\, r\right)_{s}$, upon crossing the Anti-Stokes line one has:
\bea
\left(p,\, r\right)_s \rightarrow \left(p,\, r\right)_d , \quad   \left(r,\, p\right)_d \rightarrow \left(r,\, p\right)_s
\eea
The same rule equally applies if we pick $\left(p,\, r\right)$ as dominant.  Beyond the anti-Stokes line the Airy function grows exponentially and on the line where $\text{Arg}(r)=2\pi/3$, it attains its largest value. This line is called Stokes line, on which the exponent becomes purely real. Beyond this line the form of the solution now includes a subdominant term as given in (\ref{airy}). The factor of $i$ in front of the subdominant solution is called the Stokes constant. To  see why such a change in the form  occurs, it is illustrative to analyze the function by its integral representation:
\bea
\text{Ai}(r) = \frac{1}{2\pi i} \int_{C} e^{-\frac{1}{3}t^3 + r t }\, dt
\label{airyI}
\eea
Here, the integration contour $C$  asymptotes to infinity along the directions:
\bea
|t| \rightarrow \infty  , \quad  &\frac{\pi}{2}\leq \text{Arg}(t) \leq \frac{5\pi}{6}&\nonumber\\
|t| \rightarrow  \infty , \quad  & \frac{7\pi}{6}\leq \text{Arg}(t) \leq \frac{3\pi}{2}&
\label{sect}
\eea
ensuring the finiteness of the integral given in (\ref{airyI}). One can deform $C$ to coincide with the steepest descent lines of the integrand, through single or both saddle points of the exponent given by $t_{1,\,2}=\pm |r|^{1/2}e^{\frac{i}{2}\text{Arg}(r)}$. The important point to note here is that the shape of the contour is sensitive $\text{Arg}(r)$ because the location of  the saddle points depend on $\text{Arg}(r)$.    
For instance for $\text{Arg}(r)=0$, the steepest descent  contour starts from the third quadrant as shown in the Fig. (\ref{f2}), and receives contribution from  the  saddle point of the exponent at $t_1=-r^{1/2}$, thus yielding the subdominant solution given by the first line of (\ref{airy}). As $\text{Arg}(r)$ increases, saddle points shift and beyond the Stokes line where $\text{Arg}(r)\geq 2\pi/3$ , the steepest descent contour through the lower saddle point receives contribution from the upper saddle point as well, ultimately resulting in the representation of $\text{Ai}(r)$ given by the second line of (\ref{airy}). A detailed contour analysis of Airy functions can be found in the treatise by Budden\cite{budden}.  

In functional analysis, such discontinuous changes in integral representations of special functions are ubiquitous. Recalling the asymptotic matching procedure, the Stokes phenomenon was in effect while performing the asymptotic expansion of Hypergeometric function, in passing from purely ingoing solution at the horizon to a linear combination of ingoing and outgoing solutions in the matching region.  It must be emphasized here that the resulting change in the WKB solutions upon crossing the Stokes line must be encoded in the coefficient of the subdominant solution, in the vicinity of the critical point, where the magnitude of the error in the dominant solution becomes large enough to consistently allow for such discontinuous jump.  In view of these observations, we may now give the second rule: upon crossing the Stokes line $S_{i}$, the subdominant solution changes according to:
\bea
c_1 \left(p,\, r\right)_{d} + c_2 \left(r,\, p\right)_{s} \rightarrow c_1 \left(p,\, r\right)_{d} + \left(s_{i} c_1 +  c_2 \right)\left(r,\, p\right)_{s}
\eea 
where $s_i$ is the Stokes constant associated with the Stokes line $S_{i}$. 

To briefly sum it up, the Stokes phenomenon is behind the appearance of the subdominant solution in the domain of interest.  The local form of solution can be specified in each domain bounded by Stokes and anti-Stokes lines by following these two simple rules. For this, the identification dominant and subdominant solutions is essential.  In quantum mechanical scattering problems,  the Stokes phenomenon  gives rise to the reflection phenomena\cite{meyer} and also plays a key role, in somewhat different context, in the application of resurgence theory to the perturbative expansions of quantum field theories\cite{resurge1,resurge2}.  As we will discuss shortly, the Hawking quanta responsible for the emission is encoded by the subdominant solution in the horizon region and the appearance of such solution is in fact brought by the requirement of causality.  

\subsection{Hawking Radiation}
\begin{figure*}
\includegraphics[width=10 cm,height=8cm]{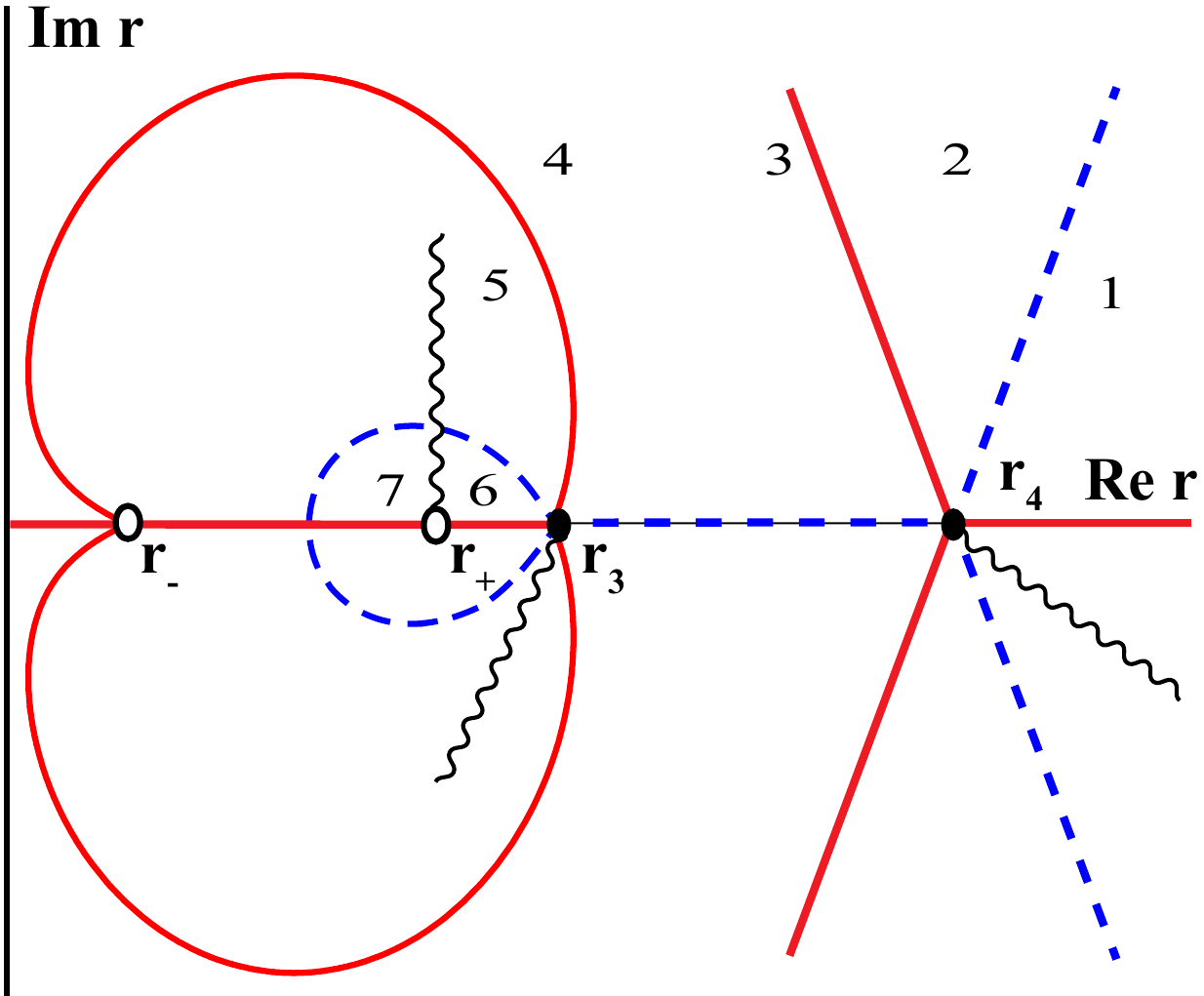}
\caption{The Stokes diagram belonging to the phase integrals $\varphi(r_3,\, r)$ and $\varphi(r_4,\, r)$. The Stokes lines (dashed, blue and where $\text{Im}\left[\varphi \right]=0 $), Anti-Stokes lines (solid,red and where $\text{Re}\left[\varphi \right]=0 $) and branch cuts (wavy, black) divide the complex plane into domains called Stokes wedges. The orientation of the cuts is in principle arbitrary, provided in any wedge the local form of the solution is maintained in accordance with the boundary conditions of the problem.}
\label{f3}
\end{figure*}

We begin the discussion by analyzing the radial momentum given in (\ref{radialm}). To make the root structure obvious, we rewrite $p_r$  by factoring out $\omega^2$ from radicand, giving:  
\bea
p_r = \dfrac{\omega \sqrt{(r-r_4) (r-r_3) (r-r_2) (r-r_1)}}{(r-r_+)(r-r_-)}
\label{radialm2}
\eea
which shows that there are four critical points and two poles, given by the location of event horizons. In the semiclassical limit ($M\omega \ll 1, \, \ell \gg 1$) all the critical points are real and satisfy: 
\bea
r_4 > r_3 > r_+ > r_- > r_2 > r_1 \quad r_{i} \in \mathbb{R},    \quad  i=1,2,3,4
\label{ordp}
\eea
and the cross ratio formed by them:
\bea
z_0=\dfrac{(r_3-r_2)(r_4-r_1)}{(r_3-r_1)(r_4-r_2)}
\eea 
satisfies $z_0 \ll 1$. The Stokes diagrams belonging to the phase integrals $ \varphi(r_4,\, r) $ and $ \varphi(r_3,\, r) $ are shown in the Fig. (\ref{f3}). The Stokes wedge 1  contains the asymptotic region $r \rightarrow \infty$, where we impose the boundary condition that there is only the outgoing wave. This is given by the solution $e^{-i\omega t}(r_4,\, r)$. To see this is indeed the case, let us resolve $\varphi (r_4,\, r)$ into real and imaginary parts:
\bea
\varphi (r_4,\, x + i y)= u (x\, ,y) + i v(x,\,y)\nonumber
\eea
At a later time $t + \delta t$, the outgoing wave on the real axis travels to the right by $x + \delta x $ such that the value of the phase becomes
\bea
i \left( v(x,\, 0) + \delta x  \dfrac{\partial v(x,\, 0)}{\partial x} \right) - i \omega (t +\delta t) \nonumber
\eea
Away from the critical point, this  is equal to the original phase provided: 
\bea
\dfrac{ d x}{d t} = \dfrac{\omega}{p_r}
\label{pv1}
\eea
Recalling we have chosen $p_r$ as the positive root from the beginning, it follows that the rhs of (\ref{pv1}) is positive as well, showing $(r_4,\, r)$ indeed represents outgoing wave, for which  $d x /d t>0$ by definition.  Note that above argument can also be given in terms of group velocity if one instead works with wave packets. The basic conclusion remains the same.

We will now perform the analytic continuation of the solution from Stokes wedge 1 to Stokes wedge 7 by making use of the rules given in the previous section. We will first assume that modes are superradiant:  $\omega- m \Omega <0$, which is generically satisfied when $M\omega \ll 1, \, m > 1$.  Above the real axis, $u (x,\, y)$ becomes negative therefore  in Stokes wedge 1 we have the subdominant solution:  
\bea
1\, :\, c_1 \left(r_4,\, r\right)_{s} 
\eea
which becomes dominant:
\bea
3\, :\, c_1 \left(r_4,\, r\right)_{d} 
\eea
in wedge 3. In order to obtain the local form of the solution near $r_3$,  the phase reference point must be changed from $r_4$ to $r_3$. As a result, the solution in region 4 is represented as:
\bea
4\, :\, c_1 \left(r_3,\, r\right)_{s} e^{S}
\eea
where the semiclassical action $S$ is given by the phase integral:
\bea
S= i \int\limits^{r_3}_{r_4} p_r \, dr ,\quad S>0
\eea
It must be noted here that an exponentially large term was factored out from the solution by changing the phase reference point and because of this  $\left(r_3,\, r\right)$ now represents the subdominant solution. This argument would of course be reversed if an exponentially small term was factored out from the subdominant solution, making it the dominant solution in the next domain. This observation, in fact, is an addendum to the analytic continuation rules of the previous section:  upon changing the phase reference point, exponentially large and exponentially small terms  appearing in the coefficients change the dominancy of the solutions\cite{heading,white}. Now proceeding to region 5 we cross the anti-Stokes line giving:
\bea
5\, :\, c_1 \left(r_3,\, r\right)_{d} e^{S}
\eea   
In passing to domain 6 one has in conjunction with the Stokes phenomenon: 
\bea
6\, :\, c_1 \left(r_3,\, r\right)_{d} e^{S} +  c_1 s \left(r,\, r_3\right)_{s} e^{S}
\label{dom6}
\eea  
Thus we have two linearly independent solutions outside the horizon. This result is in agreement with the boundary condition employed by DeWitt \cite{dewitt}. Here, $s$ is the Stokes constant which is shortly to be determined.  The identification of ingoing and outgoing solutions in wedge 6 can be done by following the argument given in \cite{bardeen}.   In the frame of the physical observer rotating with black hole, the angular velocity near the horizon is approximately given by $d\phi /dt \sim \Omega $, thus time dependent part of the solution becomes: $\sim e^{-i ( \omega-  m\Omega ) t}$. The phase velocity belonging to $ \left(r_3,\, r\right)_d$ is then:
\bea
\dfrac{d x} {dt} = \dfrac{(\omega- m\Omega)}{p_r}
\label{pv2}
\eea 
By causality, we require the solution $ \left(r_3,\, r\right)$ in wedge 7 must represent the wave with negative phase velocity and because of this we must have $p_r >0$ for $r< r_+$. This is indeed satisfied by our earlier choice of positive root for $p_r$ which led to a positive phase velocity in wedge 1. From this discussion, and minding the fact that $p_r$ reverses its sign at the horizon, it immediately follows that in wedge 6 the subdominant solution represents the ingoing solution, whereas dominant solution is identified as the outgoing wave.    

The remaining task is to trace the dominant solution in wedge 5  into the wedge 7. Note that the domains 6 and 7 are bounded by the same Stokes and anti-Stokes lines yet separated by the pole at the event horizon, where the phase integral $\varphi (r_3,\, r)$ has a branch cut discontinuity. The cut structure is complicated owing to the appearance of incomplete elliptic integral of third kind in the analytic formula for the phase integral. One may however approximately represent the discontinuity by inserting a logarithmic branch cut as this is evident from the form of near region solution given in (\ref{subs}).  The analytic continuation path is chosen to be aligned with the logarithmic branch cut, fully enclosing the pole in clockwise fashion, in order to avoid the factor two ambiguity\cite{ahmedov}. This factors out a full residue from the phase integral on the other side of the cut. Close to the pole, prefactor behaves like:  $1/\sqrt{p_r} \sim (r-r_+)^{\frac{1}{2}}$ therefore we must introduce overall minus factor to keep the solution single valued. In view of these observations we may give the solution in wedge 7 as: 
\bea
7\, :\, -c_1 \left(r_3,\, r\right)_{d} e^{S}e^{-P} 
\label{dom7}
\eea  
where the residue $P$ is:
\bea
P=\pi \sqrt{1+ \left(\frac{\omega - m \Omega}{2 \pi T_H}\right)^2}
\label{pole}
\eea
Note that the solutions in wedge 6 and 7 have approximately the same magnitude.  This is to say that the magnitude of $\left(r_3,\, r\right)_d$ in wedge 7 grows by a factor of $\sim e^{P}$ when compared to the solution in wedge 6. This behavior can be seen by analyzing the near region solution  which behaves in the upper complex plane as\cite{bcf}:
\bea
(r-r_+ +i\epsilon)^{i\frac{\omega - m \Omega}{2 \pi T_H}}=  \begin{cases}
              |(r-r_+|^{i\frac{\omega - m \Omega}{2 \pi T_H}} e^{-\frac{\omega - m \Omega}{2 T_H}}  \quad r<r_+ ,\\
              (r-r_+ )^{i\frac{\omega - m \Omega}{2 \pi T_H}} \quad r>r_+
            \end{cases}
\label{ac}
\eea   
Here, the exponent is positive for superradiant modes.  To determine the Stokes constant, we first note that $\tilde{R}$  satisfies
\bea
\text{Im}\left[\partial_r \tilde{R}\, \tilde{R}^*\right]=\text{constant}, 
\eea
which is the conserved flux along the real line. Flux conservation on both sides of the cut leads to:
\bea
|c_1|^2 e^{-2P} = |c_1|^2- |c_1|^2|s|^2
\eea 
which fixes the Stokes constant up to a phase:
\bea
s=\sqrt{1-e^{-2P}} e^{i w}
\eea
The connection formula relating domain 1 and 6 can be written as:
\bea
1\,: c_1\left(r_4,\, r\right)_{d} \,\, \rightarrow \,\,  6\, :\, c_1 \left(r_3,\, r\right)_{d} e^{S} +  c_1 \sqrt{1-e^{-2P}} e^{i w} \left(r,\, r_3\right)_{s} e^{S}
\eea
Using the above connection formula, the semiclassical decay rate is given as the modulus square of the transmission coefficient belonging to the ingoing solution at the horizon:
\bea
\Gamma_{\text{semicl}}= \frac{e^{-2S}}{1-e^{-2P}}
\label{sdecay1}
\eea
This is our main result. The total decay rate is given in terms of semiclassical greybody factor $e^{-2S}$ weighted by a Boltzmann-like factor and it remains regular in the limit $\omega \rightarrow m \Omega$. Comparing the above formula with the original Boltzmann weight given in (\ref{decay}), we see that both factors approach to unity in the low energy limit and for $m>1$. Note that the Boltzmann weight given in  (\ref{decay}) becomes negative in superradiant regime but the  total amplitude remains positive owing to the factor of $\omega- m\Omega$ coming from the greybody factor. 

For  $\omega -m \Omega >0$, analytic continuation of the modes is a little more involved. Here, we give the final formula and defer the details of the calculation to appendix. In this regime the semiclassical decay rate reads:
\bea
\Gamma_{\text{semicl}}= \frac{e^{-2S}}{e^{2P}-1}
\label{sdecay2}
\eea
which differs  from (\ref{sdecay1}) by an overall sign because the exponent now becomes positive.  This is consistent with  (\ref{decay}), where Boltzmann weight is generically close to unity (with the term $\omega-m\Omega$ factored in) for low energy modes with slow rotation ($a\ll M$ or the Schwarzschild case) or it can lead to an exponential suppression for the modes with $m<0$.  

In Hawking's original treatment using the collapse geometry and also in the analytic continuation method given by Damour and Ruffini, the decay rate is simply given by (\ref{decay}). Here, the appearance of two different formulas in two distinct regimes is the byproduct of using (\ref{schro}) along with the WKB solutions.  Recalling (\ref{pole}), the sign of the residue at $r_+$ remains uniform regardless of the sign of $\omega-m\Omega$. But the overall sign of the exponent in the Boltzmann-like factor depends on the direction of the analytic continuation path, which is chosen in accordance with the requirement of causality, and this indeed depends on the sign of $\omega-m\Omega$. To reiterate, the fact that residue does not vanish when $\omega=m\Omega$ ensures that the semiclassical decay rate remains finite. In asymptotic matching procedure on the other hand, the Boltzmann factor indeed diverges at the onset of superradiance but the total decay rate remains finite since there is a compensating factor of  $\omega- m\Omega$, coming from the conserved flux belonging to the near region solution. For WKB solutions the form of the conserved flux remains uniform, there is no compensating factor, but the Boltzmann-like factor remains finite. 

At this point  it is illustrative to contrast the method presented here with Damour and Ruffini's approach. The authors of \cite{dr} perform the analytic continuation of the outgoing solution in Kerr-Newman metric from $r>r_+$ to $r<r_+$. We will however consider the Schwarzschild case in the following  for simplicity. The outgoing solution in Schwarzschild metric reads:
\bea
\Phi_{\omega} &=& \left(2\pi |\omega|r^2 \right)^{-1/2} E^{\,\text{out}}_{\omega}(r_{*},\,t) Y^{m}_{\ell}(\theta,\, \phi) \nonumber\\
E^{\,\text{out}}_{\omega} &=& \left(r-2M \right)^{4M \omega i} e^{-i \omega (t+r_{*})}\nonumber\\
r_{*}&\sim & 2M\, \text{log}(r-2M)
\eea
Using horizon regular coordinates, above solution is mapped to an anti-particle solution for $r< 2M$. The full solution covering the beneath and beyond the horizon regions can be given by using the identity (\ref{ac}) in lower plane, yielding:
\bea
\tilde{P}= \tilde{N}_{\omega} \left[ \Theta(r-2M) \Phi^{\text{out}}_{\omega}(r-2M)+ e^{4\pi M\omega}\Theta(2M-r)\Phi^{\text{out}}_{\omega}(2M-r) \right]
\label{aps}
\eea
where $\Theta$ is the Heaviside function. Here, $\tilde{P}$ represents the anti-particle state with negative norm:
\bea
\langle \tilde{P}_{\omega_1}, \, \tilde{P}_{\omega_2} \rangle = - \delta(\omega_1 - \omega_2) \delta_{\ell_1,\, \ell_2}\delta_{m_1,\, m_2}
\eea
Equation (\ref{aps}) describes the splitting of the total state to its particle and anti-particle components.  Because $\tilde{P}$ is normalized to negative unity,  the magnitude $|\tilde{N}|^2$ of the outgoing part of the solution for $r>2M$ can be readily given as:
\bea
|\tilde{N}|^2=\frac{1}{e^{8\pi M \omega}-1}
\eea
This factor is interpreted as the particle production rate. In contrast to Damour and Ruffini's method, we have followed here the analytic continuation rules based on the exponential dominancy and subdominancy of the WKB solutions. This picture, perhaps lacks in simplicity of the method used in \cite{dr}, treats the problem as a standard quantum mechanical scattering problem without resorting to the negative norm states: the purely ingoing solution in Stokes wedge 7 undergoes over-the-barrier scattering across the horizon, and by Stokes phenomenon we have both ingoing and outgoing solutions in wedge 6.  In view of this, the particle production rate can be given as the modulus square of relative transmission coefficient in wedge 6, which is nothing but the Boltzmann-like factor. This is completely analogous to the Schwinger pair production in time-dependent laser pulses, where one has over-the-barrier scattering in a time dependent potential \cite{popov,dumlu1,dumlu2,gies,kim1}.

\subsection{Analytic Structure of the Semiclassical Penetration Factor}

In this section we treat the semiclassical action, $S$ by considering the improper integral:
\bea
S_r = i \int  \dfrac{\omega \sqrt{(r'-r_4) (r'-r_3) (r'-r_2) (r'-r_1)}}{(r'-r_+)(r'-r_-)} \, dr'
\eea
whose evaluation was performed in Mathematica\textsuperscript{\textcopyright}. After collecting and simplifying the  terms with the common denominators, we have the following expression for $S_r$ \cite{conv}:
\bea
\frac{S_r}{i \omega } &=& \dfrac{(r-r_1) (r-r_3)(r-r_4)}{\sqrt{(r-r_1)(r-r_2)(r-r_3)(r-r_4)}} -r_{21}\sqrt{\frac{r_{42}}{r_{31}}}\,\textbf{F}\left[\text{arcsin} \left[\frac{(r-r_1)r_{42}}{(r-r_2) r_{41}}\right]^{\frac{1}{2}},  \, z_0 \right]\nonumber\\
&+&  \sqrt{r_{31}{r_{42}}} \, \,
\textbf{E}\left[\text{arcsin} \left[\frac{(r-r_1)r_{42}}{(r-r_2) r_{41}}\right]^{\frac{1}{2}},  \, z_0 \right]
-  \frac{ r_{21}}{\sqrt{r_{31}r_{42}}}  f\, \boldsymbol{\Pi} \left[ \frac{r_{41}}{r_{42}} , \,\text{arcsin} \left[\frac{(r-r_1)r_{42}}{(r-r_2) r_{41}}\right]^{\frac{1}{2}},  \, z_0 \right]\nonumber\\
&+& 2\frac{ r_{21} \, r_{3-}\, r_{4-}}{r_{+\,-}\sqrt{r_{31}r_{42}}}\, \boldsymbol{\Pi} \left[z_{-} , \,\text{arcsin} \left[\frac{(r-r_1)r_{42}}{(r-r_2) r_{41}}\right]^{\frac{1}{2}},  \, z_0 \right] 
- 2\frac{ r_{21} \, r_{3+}\, r_{4+}}{r_{+\,-}\sqrt{r_{31}r_{42}}}\, \boldsymbol{\Pi} \left[z_{+} , \,\text{arcsin} \left[\frac{(r-r_1)r_{42}}{(r-r_2) r_{41}}\right]^{\frac{1}{2}},  \, z_0 \right]\nonumber\\
\eea
Here, we have used the shorthand notation: $r_{ij}:= r_i-r_j$. The functions  $\textbf{F}$, $\textbf{E}$ and $ \boldsymbol{\Pi}$  denote the incomplete elliptic integrals of first, second and third kind respectively. The factor $f$ in the fourth term above is the divisor belonging to the function $p_r$ and it is defined as:
\bea
f &=& \sum k_{z} \, \text{zeroes}-  k_{p} \, \text{poles} \nonumber\\
&=& r_1+r_2 + r_3 +r_4 - 2 (r_+ + r_-)   
\eea
where $k_z$ and $k_p$ denote the multiplicity of the zeroes and poles. The cross ratios are given as:
\bea
z_0= \frac{r_{32}\, r_{41}}{r_{31} \, r_{42}}, \,\, z_- = \frac{r_{-2}\,  r_{41}}{r_{-1}\,  r_{42}} ,\,\,  z_+ = \frac{r_{+2}\,  r_{41}}{r_{+1}\,  r_{42}},
\eea
In the limit $r\rightarrow r_4$, the angular argument becomes
\bea
\lim\limits_{r\rightarrow r_4}\text{arcsin} \left[\frac{(r-r_1)r_{42}}{(r-r_2) r_{41}}\right]^{\frac{1}{2}}=\frac{\pi}{2}
\eea
As a result, all the incomplete integrals turn into complete integrals by definition:
\bea
\frac{S_4}{i \omega } &=& -r_{21}\sqrt{\frac{r_{42}}{r_{31}}}\,\textbf{K}\left( z_0 \right)+  \sqrt{r_{31}{r_{42}}} \, \,
\textbf{E}\left( z_0 \right)
- \frac{ r_{21}}{\sqrt{r_{31}r_{42}}}  f\, \boldsymbol{\Pi} \left( \frac{r_{41}}{r_{42}} \bigg|  \, z_0 \right)\nonumber\\
&+& 2\frac{ r_{21} \, r_{3-}\, r_{4-}}{r_{+\,-}\sqrt{r_{31}r_{42}}}\, \boldsymbol{\Pi} \left(z_{-} |   \, z_0 \right) 
- 2\frac{ r_{21} \, r_{3+}\, r_{4+}}{r_{+\,-}\sqrt{r_{31}r_{42}}}\, \boldsymbol{\Pi} \left(z_{+} |   \, z_0 \right)
\label{s1}
\eea
where $\textbf{K}$ denotes complete elliptic integral of first kind. In the limit $r\rightarrow r_3$ one has:
\bea
\lim\limits_{r\rightarrow r_3}\text{arcsin} \left[\frac{(r-r_1)r_{42}}{(r-r_2) r_{41}}\right]^{\frac{1}{2}}=\text{arcsin} \left(\frac{1}{z_0}\right)^{\frac{1}{2}}=\text{arccsc} \left(\sqrt{z_0 }\right)
\eea
In this limit, the angular value hits the branch points of  $\textbf{F}$, $\textbf{E}$ and $ \boldsymbol{\Pi}$. For further simplification, we use the following relations \cite{wolfram,carl}:
\bea
\textbf{F}(\text{arccsc}(\sqrt{z_0}), z_0) &=& \frac{1}{\sqrt{z_0}}\textbf{K}\left(\frac{1}{z_0}\right) \nonumber\\
\textbf{E}(\text{arccsc}(\sqrt{z_0}), z_0) &=& \sqrt{z_0} \left( \textbf{E}\left(\frac{1}{z_0}\right)+ \left(\frac{1}{z_0}-1\right)\textbf{K}\left(\frac{1}{z_0}\right)\right)\nonumber\\
\boldsymbol{\Pi}(z,\, \text{arccsc}(\sqrt{z_0}),\,z_0) &=&\frac{1}{\sqrt{z_0}}\boldsymbol{\Pi}\left(\frac{z}{z_0},\,\frac{1}{z_0}\right)
\eea
which hold up to machine accuracy in the upper plane above the cut. Using the relations above and performing algebraic simplifications leads to:
\bea
\frac{S_3}{i \omega } &=& -r_{21}\sqrt{\frac{r_{32}}{r_{41}}}\,\textbf{K}\left(\frac{1}{z_0} \right)+  \sqrt{r_{32}{r_{41}}} \, \,
\textbf{E}\left(\frac{1}{ z_0} \right)
- \frac{r_{21}}{\sqrt{r_{41}r_{32}}} f\, \boldsymbol{\Pi} \left( \frac{r_{31}}{r_{32}} \bigg|  \, \frac{1}{z_0} \right)\nonumber\\
&+& 2\frac{ r_{21} \, r_{3-}\, r_{4-}}{r_{+\,-}\sqrt{r_{32}r_{41}}}\, \boldsymbol{\Pi} \left(\frac{z_{-}}{z_0} \bigg|   \, \frac{1}{z_0} \right) 
- 2\frac{ r_{21} \, r_{3+}\, r_{4+}}{r_{+\,-}\sqrt{r_{32}r_{41}}}\, \boldsymbol{\Pi} \left(\frac{z_{+}}{z_0} \bigg|   \, \frac{1}{z_0} \right) 
\label{s2}
\eea
Note that interchanging $r_3$ and $r_4$ yields the following relations on the cross ratios: 
\bea
z_0 \left( r_3  \leftrightarrow r_4 \right) \rightarrow \frac{1}{z_0}, \,\,  z_-\left( r_3  \leftrightarrow r_4 \right) \rightarrow \frac{z_-}{z_0} ,\,\,  z_+\left( r_3  \leftrightarrow r_4 \right) \rightarrow \frac{z_+}{z_0}
\eea
This observation readily shows that $S_3$ can be obtained from $S_4$ upon interchanging $r_3$ and $r_4$. With this in mind, the final form of the semiclassical action can be written as:
\bea
S &=& -i\omega \left(-r_{21}\sqrt{\frac{r_{42}}{r_{31}}}\,\textbf{K}\left( z_0 \right)+  \sqrt{r_{31}{r_{42}}} \, \,
\textbf{E}\left( z_0 \right)
- \frac{ r_{21}}{\sqrt{r_{31}r_{42}}}  f\, \boldsymbol{\Pi} \left( \frac{r_{41}}{r_{42}} \bigg|  \, z_0 \right) \right. \nonumber\\
&+& \left. 2\frac{ r_{21} \, r_{3-}\, r_{4-}}{r_{+\,-}\sqrt{r_{31}r_{42}}}\, \boldsymbol{\Pi} \left(z_{-} |   \, z_0 \right) 
- 2\frac{ r_{21} \, r_{3+}\, r_{4+}}{r_{+\,-}\sqrt{r_{31}r_{42}}}\, \boldsymbol{\Pi} \left(z_{+} |   \, z_0 \right) - \left(r_3 \leftrightarrow r_4 \right) \right)
\label{s3}
\eea
Here, the imaginary parts of $S_3$ and $S_4$ cancel, therefore $S$ is real and positive. One might ask at this point whether it is possible to  express (\ref{s3}) in terms of known functions. There indeed exists following relations between the complete elliptic integrals and the Hypergeometric function \cite{conv,carl}:
\bea
\textbf{K}(z)&=&\frac{\pi}{2} \mathstrut_{2}F_{1}\left(\frac{1}{2},\frac{1}{2};1;z\right)\nonumber\\
\textbf{E}(z)&=&\frac{\pi}{2} \mathstrut_{2}F_{1}\left(\frac{1}{2},-\frac{1}{2};1;z\right)\nonumber\\
\boldsymbol{\Pi}\left(z',\, z \right) &=& \frac{\pi}{2}F_{1}\left(\frac{1}{2},\frac{1}{2};1;1,z ,-z' \right)
\eea
In the last line $F_1$ represents the Appell function, a two parameter generalization of the Hypergeometric function. Unfortunately the convergence of Appell function when $z$ approaches to unity is extremely slow. For this reason we will use (\ref{s3}) in the remainder of the analysis. A much simpler expression can be obtained from (\ref{s3}) by making a power series expansion for $z_0\ll 1$. In the small cross ratio limit, we can readily observe that $S_4$'s contribution to the real part is vanishingly small because the cross ratio dependence is of the form $\sim z_0^{n}$ where $n$ is a positive integer. Thus the dominant contribution comes from $S_3$ whose cross ratio dependence is of the form $\sim 1/z_0^{n}$.  Expanding and simplifying the real part of  (\ref{s2})  in the leading order yields the following :
\bea
\text{Re} \left[ S_3 \right] &\approx& \omega \left(\frac{1}{2} \frac{r_{21}r_{41}}{\sqrt{r_{31}r_{42}}}\,\text{Log}\frac{16}{z_0}- \sqrt{r_{31} r_{42}} +2\,\text{arctan}\left[\sqrt{\frac{z_-}{z_0-z_-}} \, \right] \frac{r_{21}r_{4-}}{r_{+-}}\sqrt{\frac{r_{3-}r_{-2}}{r_{21}r_{42}}} \right. \nonumber\\ 
&-& \left.  2\,\text{arctan}\left[\sqrt{\frac{z_+}{z_0-z_+}} \, \right] \frac{r_{21}r_{4+}}{r_{+-}}\sqrt{\frac{r_{3+}r_{+2}}{r_{21}r_{42}}}\, \right)
\label{s4}
\eea 
One may further simplify this expression by noting ($\ell \gg 1$):
\bea
-r_1\approx r_4 \approx \dfrac{\sqrt{\lambda}}{\omega} \gg  \left (r_{3},\, r_{2}\right)
\label{r1r4}
\eea
which leads to:
\bea
\text{Re} \left[ S_3 \right] &\approx& \left( \lambda^{1/2} \, \text{Log}\frac{16}{z_0}-\lambda^{1/2} +2 \,\lambda^{1/2}\frac{\sqrt{r_{3-}r_{-2}}}{r_{+-}} \,\text{arctan}\left[\sqrt{\frac{z_-}{z_0-z_-}} \, \right]  \right. \nonumber\\ 
&-& \left.  2\, \lambda^{1/2}\frac{\sqrt{r_{3+}r_{+2}}}{r_{+-}}\, \text{arctan}\left[\sqrt{\frac{z_+}{z_0-z_+}} \, \right]\ \right)
\label{s5}
\eea
Here, $z_{\pm}$ appear as angular variables whose interpretation remains somewhat elusive. We should however note that a closer inspection using (\ref{r1r4}) shows that last two terms above multiplying the angular terms resemble residues of the action at $r_{\pm}$. For instance at $r_{+}$, the residue of $S_r$ can be written as:
\bea
\text{Res}\left[S_{r}\right] &=& i\,\omega\frac{\sqrt{r_{+4}\,r_{+3}\,r_{+2}\,r_{+1}}}{r_{+-}} \nonumber\\
&\approx & i\,\lambda^{1/2} \frac{\sqrt{r_{3+}\, r_{+2}}}{r_{+-}}
\eea 
Thus for large $\ell$ the associated Cauchy integral (in the positive sense) is:
\bea
S_{r_+} \approx  -2 \,\lambda^{1/2} \frac{\sqrt{r_{3+}r_{+2}}}{r_{+-}} \pi 
\eea 
which is identical to the last term in (\ref{s5}) with $\text{arctan}$ function replaced by a factor of $\pi$.  It is instructive to see at this point how the power law dependence of the greybody factor $\sigma \sim (r_+-r_-)^{2\ell +1} \omega^{2\ell +1}$ follows from (\ref{s5}). For very large $\ell$, the critical points $r_3$ and $r_2$ slowly approach to $r_+$ and $r_-$ respectively. In this limit, the first two terms in (\ref{s5}) dominate so we may write:
\bea
S\approx  \lambda^{1/2} \left(\text{Log}\frac{16}{z_0}-1 \right)
\label{s6}
\eea 
and the cross ratio becomes approximately: 
\bea
z_0 \approx \frac{2\,\omega (r_+-r_-)}{\lambda^{1/2}}
\label{cr}
\eea
Using (\ref{s6}) and (\ref{cr}) yields the barrier penetration factor in the leading order:
\bea
e^{-2S} \approx  \frac{\omega^{2\ell +1} \left(r_+-r_- \right)^{2\ell +1}}{2^{2\ell+1}} \frac{e^{2\ell+1}}{\left(4\ell +2\right)^{2\ell +1}}
\eea
where we have used $\lambda^{1/2}\approx \ell+\frac{1}{2}$. This shows that semiclassical decay rate given by (\ref{sdecay1}) precisely shows the same power law behavior as expected. For a more precise comparison between $\Gamma_{\text{semicl}}$ and $\Gamma$   we will in the following numerically compare $e^{-2S}$ and $\sigma$, using the full expressions given by (\ref{s3}) and (\ref{abs}).  But since these expressions get damped very quickly as $\ell$ gets large,  for convenience we define log-scale difference:
\bea
\Delta_e = \big|\text{Log}\,\sigma +2S\big|, \quad \text{Log}\,\sigma <0
\eea
\begin{figure}[t!]
\includegraphics[width=5.4 cm,height=4cm]{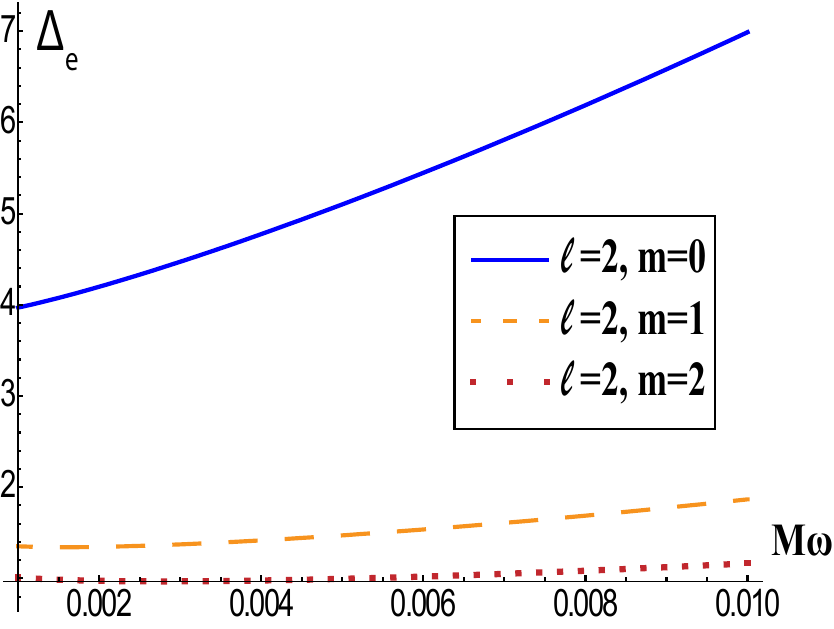}
\includegraphics[width=5.4 cm,height=4cm]{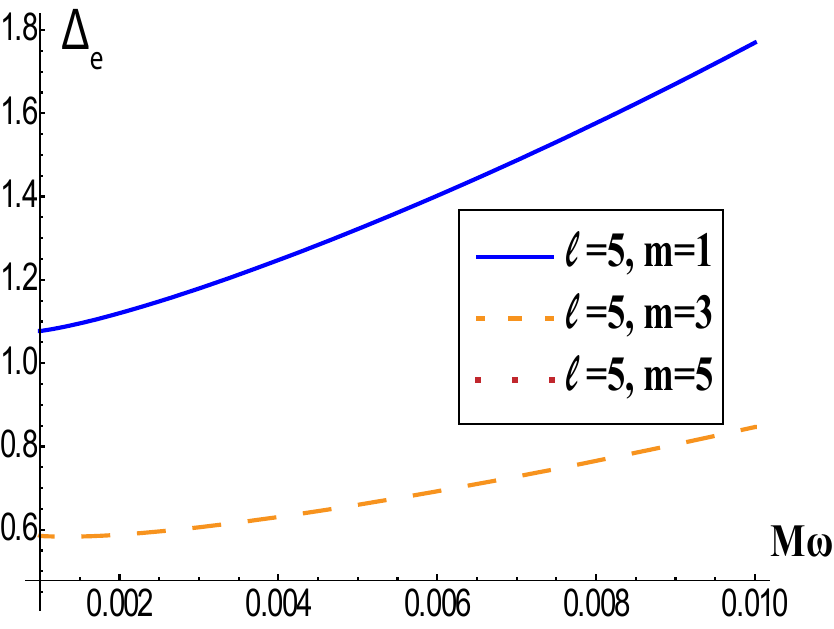}
\includegraphics[width=5.4 cm,height=4cm]{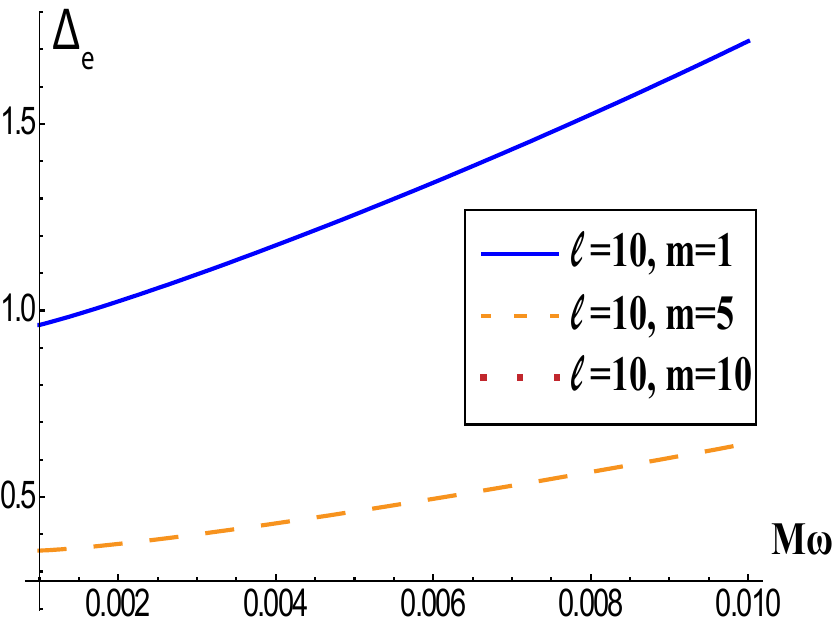}
\caption{The numerical comparison between asymptotic matching formula given in (\ref{abs}) and the leading order semiclassical penetration factor given by (\ref{s3}). The relative error diminishes as $\ell$ gets large, for the fixed value of $\ell$ the relative error assumes its minimum when $m=\ell$.  }
\label{f4}
\end{figure}We evaluate the above expression for low energy superradiant modes, for which Boltzmann factors can be taken as unity.  Figure (\ref{f4}) shows the numerical data for $\Delta_e$ for various values of $\ell$ and $m$. For fixed $\ell$, the relative error is the largest for $m=0$ and diminishes for increasing values of $m$.  As $\ell$ gets larger, the relative error accordingly drops down further.   As mentioned in the introduction, the accuracy of the leading order semiclassical formula is limited, giving reasonably accurate results with the asymptotic matching formula when the condition: $M\omega \ll 1,\, \ell \gg 1$ is satisfied. For high energy modes where $M \omega >1$, the use of numerical techniques or higher order WKB action becomes essential for more accurate results \cite{kanti,rocha,konoplya}. It is worth noting here that at high energies the critical points move to complex plane in the form of complex conjugate pairs. All the cross ratios become complex, specifically $z_0$ being real grows up to $\sim 1$ first, then assumes the form $z_0= e^{i \rho}$ where $\rho$ is a real number.

\section{Conclusion}

In our analysis we have incorporated Stokes phenomenon to the tunneling problem of massless scalar field from fixed Kerr background. In doing so, we have employed standard analytic continuation rules based on exponential dominancy of the WKB solutions. With the one way connection formula that relates the asymptotic solution to  the near horizon solutions, we were able obtain compact expressions for the semiclassical decay rate. We have given the semiclassical penetration factor in terms of complete elliptic integrals, whose arguments depend on three independent cross ratios, $(z_0,\, z_{\pm})$. We have observed that the logarithmic dependence of the action on $z_0$ leads to the power law structure,  a property which is already manifest in the asymptotic matching formulas.  Interestingly enough, the remaining cross-ratios  $z_{\pm}$  appear as angular variables in the action, resembling the multiplicative angular factors of Cauchy integrals evaluated at $r_{\pm}$.

\section{Appendix}

\begin{figure}[b!]
\includegraphics[width=10 cm,height=8cm]{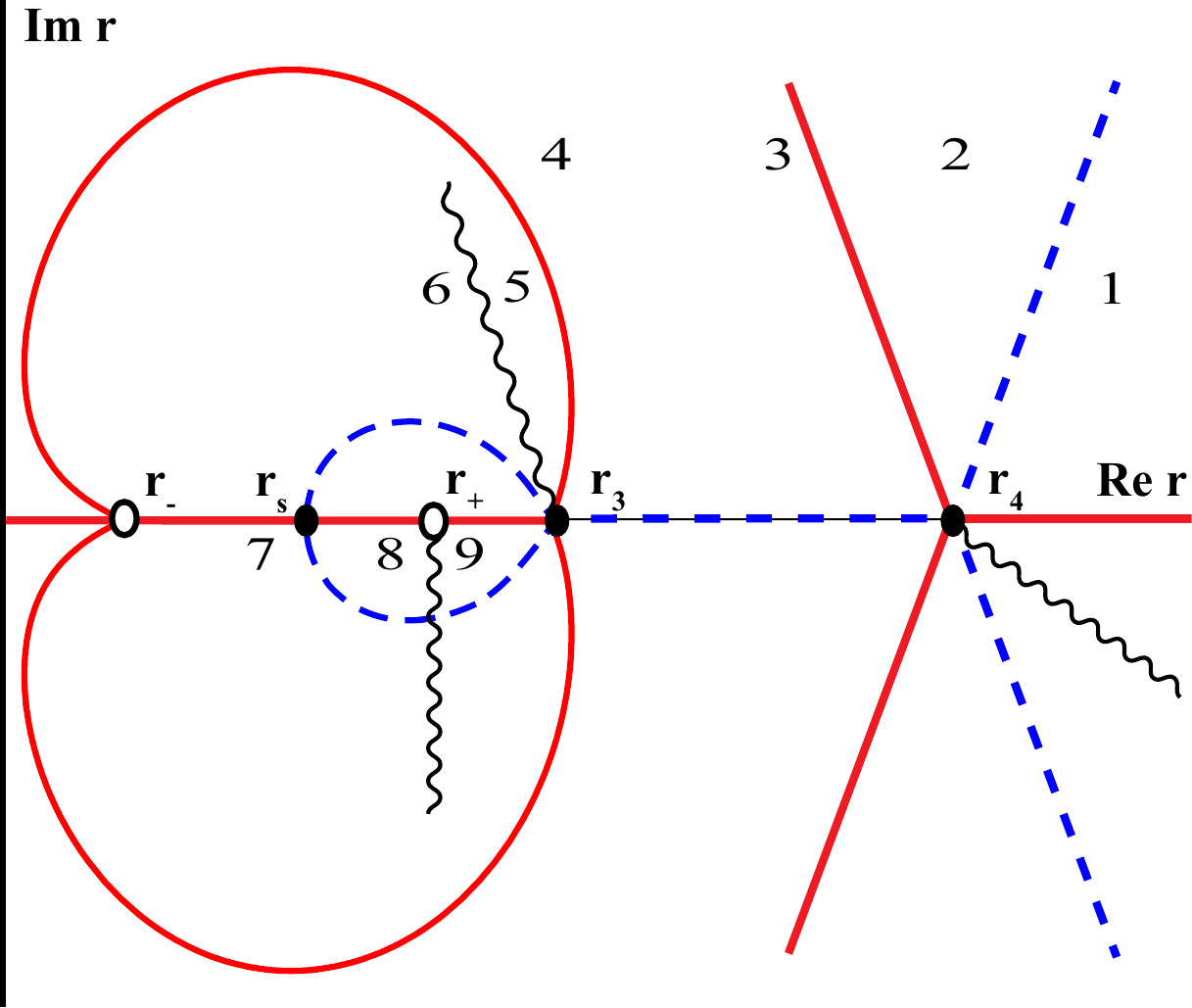}
\caption{The Stokes diagram belonging to the phase integrals $\varphi(r_3,\, r)$ and $\varphi(r_4,\, r)$ for $\omega - m \Omega >0$. As before, the Stokes lines (dashed, blue and where $\text{Im}\left[\varphi \right]=0 $), Anti-Stokes lines (solid,red and where $\text{Re}\left[\varphi \right]=0 $) and branch cuts (wavy, black) divide the complex plane into the depicted domains but now the orientation of the cuts emanating from $r_3$ and $r_+$ is reversed so as to satisfy the causality requirement.}
\label{f5}
\end{figure}

Here, we give the details of analytic continuation of the modes for $\omega -m\Omega>0$. Recalling (\ref{pv2}), the first thing to note here is that by the causality requirement the ingoing wave in wedge 6 must now be of the form $(r,\, r_3)$. Because of this, we place the branch cut of the critical point in upper plane and the logarithmic cut in lower plane as shown in Fig. (\ref{f5}).
The steps of analytic  continuation until the wedge 6 are identical to the ones given in section III. Recalling that in domain 5 we have the solution:
\bea
5\, :\, c_1 \left(r_3,\, r\right)_{d} e^{S}
\eea   
we need to specify the form of the solution on the other side of the branch cut.  To do so, observe that near a zero $r_0$, a generic potential $V(r)$ behaves linearly  therefore the value of $\varphi(r_0, \, r)$  differs by a factor of $e^{-3\pi i}$ on the other side of the branch cut, where $r\rightarrow |r|e^{i (\text{Arg}(r)-2\pi)}$. The square root in the prefactor on the other hand acquires a factor of $e^{-i\frac{\pi }{2}}$ (the sign of the exponents here must be reversed if the cut is traversed in the negative(clockwise) sense). To keep the solutions single-valued across the cut, these phase changes must be accounted for. This leads us to the following rule: upon crossing a branch cut in the positive  sense, the solutions must be modified as \cite{heading}: 
\bea
\left(r_0,\, r\right) \rightarrow -i\left(r,\,r_{0}\right) , \quad   \left(r,\, r_{0}\right) \rightarrow -i\left(r_{0},\, r\right)
\label{cut}
\eea
This process leaves the value of the solution of across the cut intact and as a result the exponential dominancy and subdominancy of the solutions remain the same.  Following (\ref{cut}), in domain 6 we have:
\bea
6\, :\, -i c_1 \left(r,\, r_3\right)_{d} e^{S}
\eea  
We will now perform the continuation of the above solution to the lower complex plane. Crossing the anti-Stokes line on the real axis, in domain 7 the solution becomes:
\bea
7\, :\, -i c_1 \left(r,\, r_3\right)_{s} e^{S} = -i c_1 \left(r,\, r_s\right)_{d} e^{S} e^{-P/2}
\eea  
Here, we  have carried the phase reference point to the other side of the cut from $r_3$ to $r_s$, the point where the Stokes line cuts the real axis.  This factors out a half residue from the phase integral, changing the solution from a subdominant to a dominant one in the process. 

The remaining task is to trace the solution in wedge 7 into the wedge 8 and wedge 9. By causality we require to have only the dominant solution in wedge 8:
\bea
8\, :\,  -i c_1 \left(r,\, r_s\right)_{d} e^{S} e^{-P/2}
\eea  
which is also consistent  with the fact that the Stokes jump is expected to occur in the vicinity of the critical point. To get the form of solution in wedge 9, we trace the the dominant solution along a  path aligned with the logarithmic cut, fully enclosing the pole in clockwise fashion. Taking into account extra minus factor coming from the prefactor on the other side of the cut, and the Stokes jump one ultimately has:
\bea
9\, :\,  i c_1 \left(r,\, r_s\right)_{d} e^{S} e^{P/2}   -i c_1 s \left(r_s,\, r\right)_{s} e^{S} e^{-P/2}
\eea 
As done before, the stokes constant can be determined up to a phase by  equating the fluxes belonging to domains 8 and 9, yielding:
\bea
s=\sqrt{e^{2P}-1} e^{i \tilde{w}}
\eea 
To write the connection formula in its proper form, as a final step, we carry the phase reference point back to $r_3$ along the semicircle around the pole in the lower plane. Doing so, the connection formula reads: 
\bea
1\,: c_1\left(r_4,\, r\right)_{d} \,\, \rightarrow \,\,  9\, :\, i c_1 \left(r,\, r_3\right)_{d} e^{S}   -i c_1 \sqrt{e^{2P}-1} e^{i \tilde{w}} \left(r_3,\, r\right)_{s} e^{S} 
\eea
yielding the semiclassical decay rate:
\bea
\Gamma_{\text{semicl}}= \frac{e^{-2S}}{e^{2P}-1}
\eea

\end{document}